\documentclass[a4paper,12pt]{article}
\usepackage{amsfonts}
\usepackage{latexsym}
\usepackage{amssymb}
\date{}

\begin{document}

\title{Exploitation of a special-relativistic entropy identity\\
for one component of a multi-component system
}
\author{W. Muschik\footnote{Corresponding author:
muschik@physik.tu-berlin.de}
\\
Institut f\"ur Theoretische Physik\\
Technische Universit\"at Berlin\\
Hardenbergstr. 36\\D-10623 BERLIN,  Germany}
\maketitle

            \newcommand{\be}{\begin{equation}}
            \newcommand{\beg}[1]{\begin{equation}\label{#1}}
            \newcommand{\ee}{\end{equation}\normalsize}
            \newcommand{\bee}[1]{\begin{equation}\label{#1}}
            \newcommand{\bey}{\begin{eqnarray}}
            \newcommand{\byy}[1]{\begin{eqnarray}\label{#1}}
            \newcommand{\eey}{\end{eqnarray}\normalsize}
            \newcommand{\beo}{\begin{eqnarray}\normalsize}
            \newcommand{\R}[1]{(\ref{#1})}
            \newcommand{\C}[1]{\cite{#1}}

            \newcommand{\mvec}[1]{\mbox{\boldmath{$#1$}}}
            \newcommand{\x}{(\!\mvec{x}, t)}
            \newcommand{\m}{\mvec{m}}
            \newcommand{\F}{{\cal F}}
            \newcommand{\n}{\mvec{n}}
            \newcommand{\argm}{(\m ,\mvec{x}, t)}
            \newcommand{\argn}{(\n ,\mvec{x}, t)}
            \newcommand{\T}[1]{\widetilde{#1}}
            \newcommand{\U}[1]{\underline{#1}}
            \newcommand{\V}[1]{\overline{#1}}
            \newcommand{\X}{\!\mvec{X} (\cdot)}
            \newcommand{\cd}{(\cdot)}
            \newcommand{\Q}{\mbox{\bf Q}}
            \newcommand{\p}{\partial_t}
            \newcommand{\z}{\!\mvec{z}}
            \newcommand{\bu}{\!\mvec{u}}
            \newcommand{\rr}{\!\mvec{r}}
            \newcommand{\w}{\!\mvec{w}}
            \newcommand{\g}{\!\mvec{g}}
            \newcommand{\D}{I\!\!D}
            \newcommand{\se}[1]{_{\mvec{;}#1}}
            \newcommand{\sek}[1]{_{\mvec{;}#1]}}            
            \newcommand{\seb}[1]{_{\mvec{;}#1)}}            
            \newcommand{\ko}[1]{_{\mvec{,}#1}}
            \newcommand{\ab}[1]{_{\mvec{|}#1}}
            \newcommand{\abb}[1]{_{\mvec{||}#1}}
            \newcommand{\td}{{^{\bullet}}}
            \newcommand{\eq}{{_{eq}}}
            \newcommand{\eqo}{{^{eq}}}
            \newcommand{\f}{\varphi}
            \newcommand{\rh}{\varrho}
            \newcommand{\dm}{\diamond\!}
            \newcommand{\seq}{\stackrel{_\bullet}{=}}
            \newcommand{\st}[2]{\stackrel{_#1}{#2}}
            \newcommand{\om}{\Omega}
            \newcommand{\emp}{\emptyset}
            \newcommand{\bt}{\bowtie}
            \newcommand{\btu}{\boxdot}
\newcommand{\Section}[1]{\section{\mbox{}\hspace{-.6cm}.\hspace{.4cm}#1}}
\newcommand{\Subsection}[1]{\subsection{\mbox{}\hspace{-.6cm}.\hspace{.4cm}
\em #1}}

\newcommand{\const}{\textit{const.}}
\newcommand{\vect}[1]{\underline{\ensuremath{#1}}}  
\newcommand{\abl}[2]{\ensuremath{\frac{\partial #1}{\partial #2}}}

\begin{center}
{\large Guestpaper THERMOCON2016}
\end{center}


\vspace{1cm}\noindent
{\bf Abstract} Non-equilibrium and equilibrium thermodynamics of an interacting component of a
special-relativistic multi-component system is discussed by using an entropy identity. The special
case of the corresponding free component is considered.

\section{Introduction}

The treatment of multi-component systems is often restricted to transport phenomena in
chemically reacting systems, that means, the mixture consisting of different components is
described by 1-component quantities such as temperature, pressure and energy which are not
retraced to corresponding quantities of the several components. That is the case in non-relativistic
physics \C{dGM} as well as in relativistic physics \C{K1,K2,IS,HASW,NEU}. In this paper, the single
component as an interacting member of the mixture is investigated. Thus, each component of the
mixture is equipped with its own temperature, pressure, energy and mass density which all together
generate the corresponding quantities of the mixture.
\vspace{.3cm}\newline
Considering a multi-component system,  three items have to be distinguished: a component
of the system which interacts with all other components of the system, the same component as a
free 1-component system separated from the multi-component system and finally the
multi-component system itself as a mixture which is composed of components. Here, the
interacting and the free component are discussed, both in a special-relativistic framework. For
finding out the entropy-flux, -supply, -production and -density, a special tool is used: the entropy
identity which constrains the possibility of an arbitrary choice of these quantities \C{BOCH,MUBO,
MUBO1}. The specific
entropy and the corresponding Gibbs and Gibbs-Duhem equations are derived. By use of the entropy
identity, accessory variables are introduced belonging to the balance equations which are taken
into account in the entropy identity: those of the energy-momentum tensor, the rest mass density
and the diffusion flux.  
Equilibrium is defined by equilibrium conditions which are divided into basic ones given by
vanishing entropy-flux, -supply and -production and into supplementary ones such as vanishing
diffusion flux, vanishing heat flux and rest mass production. The Killing relation of the 4-temperature
is shortly discussed. Constitutive equations are out of scope of this paper. 
\vspace{.3cm}\newline
The paper is organized as follows: After this introduction, the kinematics of a multi-component 
system is considered for introducing the mass flux density and the diffusion flux. The
energy-momentum tensor is decomposed
into its (3+1)-split, and the non-equilibrium thermodynamics of an interacting component in the
mixture and those of the corresponding free component is discussed. The equilibrium belonging
to both kinds of components finishes the paper together with a summary.

\section{Kinematics}
\subsection{The components}

We consider a multi-component system consisting of $Z$ components.
The component index $^A$ runs from $1$ to $Z$. Each component has
its own rest frame ${\cal B}^A$ in which the rest mass density $\varrho^A$
is locally defined. These rest mass densities are relativistic invariants,
that means the $\varrho^A$ are scalars. 
\vspace{.3cm}\newline
In general, the components have different 4-velocities: $u^A_k,\ A=1,2,...,Z$
which all are tensors of first order. We now define the mass flux density 
\bee{K2}
N^A_k\ :=\ \varrho^A u^A_k,\qquad N^{Ak}{_{,k}}\ =\ \Gamma^A.
\ee
Here, \R{K2}$_2$ is the mass balance equation of the $^A$-component .
Consequently, we introduce the basic fields of the components
\bee{K3}
\{\varrho^A,\ u^A_k\},\qquad A=1,2,...,Z.
\ee

\subsection{The mixture}

As each component, also the multi-component system has a mass density
$\varrho$ and a 4-velocity $u_k$ which are determined by the partial
quantities of the components. For deriving $\varrho$ and $u_k$, we apply
the
\vspace{.3cm}\newline
$\blacksquare$\ {\sf Mixture Axiom:} The balance equation of a mixture
looks like the balance equation of an one-component system.
\hfill$\blacksquare$
\vspace{.3cm}\newline
According to the mixture axiom, the mass balance of the mixture looks like  
\bee{K4}
N^k{_{,k}}\ =\ \Gamma,\qquad\Gamma\ =\ 0,
\ee
with vanishing production, if the mass of the mixture is conserved. 
\vspace{.3cm}\newline
Now the question arises: which
quantities of the components are additive ? Obviously, neither the mass
densities $\varrho^A$ nor the component 4-velocities $u^A_k$ are additive
quantities according to their definitions.
Consequently, we demand in accordance with the mixture axiom
that the mass flux densities are additive\footnote{The sign $\st{\td}{=}$ stands
for a setting and $:=$ for a definition.}
\bey\nonumber
\mbox{\sf Setting I:}\hspace{9.9cm}
\\ \label{K5}
\sum_A N^A_k\ \st{\td}{=}\ N_k\ :=\  \varrho u_k\ =\
\sum_A \varrho^A u^A_k\quad\longrightarrow\quad
u_k\ =\ \sum_A\frac{\varrho^A}{\varrho}u^A_k.
\eey
For the present, $\varrho$ and $u_k$ are unknown. Of course, they depend
on the basic fields of the components \R{K3}.
Contraction with $u^k$ and use of \R{K5}$_{2,3}$ results in
\bee{K6}
\varrho\ =\ \frac{1}{c^2}\sum_A\varrho^A u^A_ku^k\ =\ \frac{1}{c^2}N_ku^k\
=\ \frac{1}{c^2}N_k\frac{1}{\varrho}N^k\ \longrightarrow\
\varrho\ =\ \pm\frac{1}{c}\sqrt{N_kN^k},
\ee
or in more detail
\bee{K6a}
\varrho\ =\ \pm\frac{1}{c}\sqrt{\sum_{A,B}\varrho ^A\varrho^Bu^A_ku^{Bk}}.
\ee
The mass density $\varrho$ and the 4-velocity $u_k$ of the mixture are
expressed by those of the components according to \R{K6a} and \R{K5}$_4$.
According to \R{K5}$_4$, the 4-velocity of the mixture is a weighted mean value
of the 4-velocities of the components. For the mass density, we have according to
\R{K6}$_1$ also a weighted mean value of the mass density components
\bee{K6b}
f^A\ := \frac{1}{c^2}u^A_ku^k\quad\longrightarrow\quad
\rh\ =\ \sum_Af^A\rh^A\ =\ \sum_Af^A(u^A_k,u^k)\varrho^A,
\ee
resulting in the entanglement of $\rh$ and $u_k$ which are not independent of each
other
\bee{K6c}
\rh\ =\ R(\rh^A,u^A_k,u_k),\qquad u_k\ =\ U_k(\rh^A,u^A_k,\rh).
\vspace{.3cm}\ee
According to \R{K5}$_1$ and \R{K2}$_2$, we obtain the additivity of the mass
production terms
\bee{K7a}
N^k{_{,k}}\ =\ \sum_AN^{Ak}{_{,k}}\ =\ \sum_A\Gamma ^A\ =\ \Gamma.
\ee

\subsection{The diffusion flux}

From \R{K5}$_3$ and \R{K6b}$_2$ follows
\bee{M1}
0\ =\ \sum_A\rh^Au^A_k - u_k\sum_Af^A\rh^A\ =\ \sum_A \rh^A(u^A_k-f^Au_k).
\ee
Introducing the diffusion flux density using \R{M1}$_2$
\bee{K8c}
J^A_k\ :=\ \varrho^A(u^A_k - f^Au_k)\ =\ N^A_k - \rh^Af^Au_k
\quad\longrightarrow\quad
\sum_A J^A_k\ =:\ J_k\ =\ 0,
\ee
we obtain 
\byy{K9}
J^A_ku^k &=& \varrho^A(u^A_ku^k - f^Ac^2)\ =\ 0,
\\ \label{K9a}
J^A_ku^{Ak}&=& c^2\rh^A[1-(f^A)^2]\ =:\ c^2\rh^Aw^A\
=w^AN^A_ku^{Ak}
\eey
By introducing the projectors
\bee{K15}
h^{Am}_l\ :=\ \delta^m_l - \frac{1}{c^2}u^{Am}u^{A}_l,\qquad
h^{m}_l\ :=\ \delta^m_l - \frac{1}{c^2}u^{m}u_l,
\ee
we obtain the following properties of the diffusion flux density:
\byy{L1}
J^{Am}h^k_m &=& J^{Ak}\ =\ N^{Am} h^k_m
\\ \label{L3}
J^{Ak} &=& J^{Am}h^{Ak}_m + \rh^Aw^Au^{Ak}\ =\
J^{Am}h^{Ak}_m+w^AN^{Ak}                                                 
\\ \label{K8d1}
J^{Ak}{_{,k}} &=&(J^{Am}h^{Ak}_m)_{,k}
+(\rh^Aw^A){_{,k}}u^{Ak}+\rh^Aw^Au^{Ak}{_{,k}}.
\eey
According to \R{L1}$_2$, the diffusion flux density is that part of the mass flux density
which is perpendicular to the 4-velocity of the mixture. The diffusion flux density
vanishes in 1-component systems ($u^A_k\equiv u_k$) according to $f^A=f=1$ and \R{K8c}$_1$.

\section{The Energy-Momentum Tensor}
\subsection{Free and interacting components\label{FIC}}

The energy-momentum tensor $T^{Akl}$ of the $^A$-component consists of two parts
\bee{O1}
T^{Akl}\ =\ \st{0}{T}\!{^{Akl}} + \sum_B W^{Akl}_B,\quad W^{Bkl}_B\ =\ 0.
\ee
Here, $\st{0}{T}\!{^{Akl}}$ is the energy-momentum tensor of the free
$^A$-component, that is the case, if there are no interactions between the
$^A$-component and the other ones. $W^{Akl}_B$ describes the interaction
between the $^B$- and the $^A$-component.
The interaction between the environment and the $^A$-component is given by the
force density $k^{Al}$ which appears in the energy-momentum balance equation\footnote{The
force density $k^{Al}$ does not contain any interaction between the components, but only those
between the $^A$-component and the environment outside the mixture. The interaction between
the components is according to \R{O1} included in $T^{Akl}$.} 
\bee{O2}
T^{Akl}{_{,k}}\ =\ k^{Al},
\ee
and in the balance equations of
\byy{K13}
\mbox{energy:}\hspace{1.4cm}
u^A_lT^{Akl}{_{,k}} &=& u^A_lk^{Al},
\\ \label{K14}
\mbox{momentum:}\hspace{.5cm}
h^{Am}_lT^{Akl}{_{,k}} &=& h^{Am}_lk^{Al}.
\eey
Consequently, the interaction of the $^A$-component with the other components
of the mixture modifies the energy-momentum tensor of the free $^A$-component.
Additionally, its interaction with the environment shows up in the source of
the energy-momentum balance. According to its definition, $T^{Akl}$ is the energy-momentum
tensor of the "$^A$-component in the mixture".

\subsection{(3+1)-split}

The (3+1)-split of the energy-momentum tensor of the $^A$-component is
\bee{J1}
T^{Akl}\ =\ \frac{1}{c^2}e^Au^{Ak}u^{Al} + 
\frac{1}{c}u^{Ak}p^{Al} +\frac{1}{c^2}q^{Ak}u^{Al} +t^{Akl}.
\ee
The (3+1)-components of the energy-momentum tensor are\footnote{the (3+1)-split
is made by taking the physical meaning of \R{J2} and \R{J3} into account, see \R{J6} to \R{J8}}
\byy{J2}
e^A\ :=\ \frac{1}{c^2}T^{Ajm}u^A_ju^A_m ,\qquad
p^{Al}\ :=\ \frac{1}{c}h^{Al}_mT^{Ajm}u^A_j,
\\ \label{J3}
q^{Ak}\ :=\ h^{Ak}_jT^{Ajm}u^A_m,\qquad
t^{Akl}\ :=\ h^{Ak}_jT^{Ajm}h^{Al}_m,
\\ \label{J3a}
q^{Ak}u^A_k = 0,\ p^{Al}u^A_l = 0,\quad t^{Akl}u^A_k =0,\ t^{Akl}u^A_l = 0.
\vspace{.3cm}\eey
If the stress tensor is decomposed into pressure $p^A$ and viscous tensor
$\pi^{Akl}$
\bee{cT2}
t^{Akl}\ =\ -p^Ah^{Akl} + \pi^{Akl},
\ee 
we obtain for the physical dimension of the pressure\footnote{the bracket [$\boxtimes$]
signifies the physical dimension of $\boxtimes$}
\bee{J4}
[p]\ =\ \frac{N}{m^2}\ =\ \frac{Nm}{m^3}\ =\ \mbox{energy density}. 
\ee
According to \R{K15} and \R{K6b}$_1$, we have
\bee{J5}
[h^{Al}_m]\ =\ 1,\qquad [f^A]\ =\ 1,
\ee
and by taking \R{cT2}, \R{J1} and \R{J5} into account we obtain
\byy{J6}
[\pi^{Akl}]\ =\ [t^{Akl}]\ =\ [T^{Akl}]\ =\ [e^A]\ =\ [p^{Al}]\ =\ \frac{Nm}{m^3}
\ =\ \mbox{energy density}
\\ \label{J7}
[p^{A}]\ =\ \frac{N}{m^2}\ =\ \frac{kg\ m}{s^2}\frac{1}{m^2} =
kg\frac{m}{s}\frac{1}{m^3}\frac{m}{s}\ =\ \mbox{momentum flux density},
\\ \label{J8}
[q^{Ak}]\ =\ [T^{Akl}]\frac{m}{s}\ =\ \frac{Nm}{m^3}\frac{m}{s}\ =\
\mbox{energy flux density}.
\eey

\section{Thermodynamics (InteractingComponent)}
\subsection{The entropy identity}

For establishing the entropy balance equation, we use a special procedure
starting out with an identity \C{MUBO}, the so-called {\em entropy identity}.
This tool helps to restrict arbitrariness for defining entropy density, entropy flux density,
entropy production and supply. In the sequel, we establish the entropy identity 
for the $^A$-component.
Starting out with the (3+1)-split of the entropy 4-vector
\bee{T2}
S^{Ak}\ =\ s^Au^{Ak} + s^{Ak},
\ee
we lay down $s^A$ as entropy density and $s^{Ak}$ as entropy flux density
\byy{T2z}
s^A\ :=\ \frac{1}{c^2}S^{Ak}u^A_k,&\quad&
s^{Ak}\ :=\ S^{Am}h^{Ak}_m, 
\\ \label{T2w}
[s^A]\ =\ [e^A]\frac{1}{K}\ =\ \frac{Nm}{m^3}\frac{1}{K},&\quad&
[s^{Ak}]\ =\ [q^{Ak}]\frac{1}{K}\ =\
 \frac{Nm}{m^3}\frac{m}{s}\frac{1}{K}.
\vspace{.3cm}\eey
Before writing down the entropy identity, we have to choose the quantities which are
essential for formulating the four entropy quantities mentioned above. The choice is:
all quantities appearing in the (3+1)-split of the energy-momentum tensor \R{J1} and
additional the mass flux density \R{K2}$_1$ and the diffusion flux density \R{L3}$_2$ have to be included because we are
dealing with a multi-component system. Consequently, the entropy identity is
\bey\nonumber
S^{Ak}&\equiv& s^Au^{Ak} + s^{Ak}
+\ \chi^A \Big(J^{Ak}-J^{Am}h_m^{Ak}-w^AN^{Ak}\Big)
+\kappa^A\Big(N^{Ak}-\rh^Au^{Ak}\Big)
\\ \label{aT2}
&&+\ \Lambda ^A_l\Big(T^{Akl}-\frac{1}{c^2}e^Au^{Ak}u^{Al} - 
\frac{1}{c}u^{Ak}p^{Al} -\frac{1}{c^2}q^{Ak}u^{Al} -t^{Akl}\Big).
\eey
Here $\chi^A$ and $\kappa^A$ are scalars, undefined for the present, and
the (3+1)-split of the likewise arbitrary vector $\Lambda^A_l$ is
\byy{bT2}
\Lambda^A_{l}\ =\ \lambda^Au^A_{l}+ \lambda^A_l,\qquad
\lambda^A_lu^{Al}\ =\ 0,&\quad& \lambda^A_j h^{Aj}_l\ =\ \lambda^A_l,
\\ \label{bT2a}
\Lambda^A_lu^{Al}\ =\ c^2\lambda^A,&\quad&\Lambda^A_lh^{Al}_m\ =\
\lambda^A_m. 
\eey
We denote $\chi^A$, $\kappa^A$, $\lambda^A$ and $\lambda^A_l$ as "accessory variables" because
they help to formulate the entropy identity. An identification of these auxiliary variables
is given below after the definitions of entropy flux density, entropy production, density
and supply in section \ref{ACVA}.
By use of \R{cT2} and \R{bT2}, the entropy identity \R{aT2} becomes
\bey\nonumber
S^{Ak}\ \equiv\ 
\ u^{Ak}\Big(s^A
-\chi^A\rh^Aw^A
-\kappa^A\rh^A
-\lambda^Ae^A
-\frac{1}{c}\lambda^A_lp^{Al}\Big)+\kappa^AN^{Ak}
\hspace{2.3cm} 
\\ \nonumber
+\ s^{Ak} 
+\chi^AJ^{Ak} 
-\chi^AJ^{Am}h_m^{Ak}
-\lambda^Aq^{Ak} 
+\lambda^{Ak}p^A-\lambda^A_l\pi^{Akl}+
\hspace{.2cm}
\\ \label{dT2}
+\ \Big(\lambda^Au^A_{l}+ \lambda ^A_l\Big)T^{Akl}.
\hspace{.3cm}
\eey
This identity becomes an other one by differentiation and by taking
the entropy balance equation
\bee{T5}
S^{Ak}{_{,k}}\ =\ \sigma^A + \varphi^A 
\ee
into account                                                                                     
\bey\nonumber
S^{Ak}{_{,k}}\ \equiv\ \Big[u^{Ak}
\Big(s^A 
-\chi^A\rh^Aw^A
-\kappa^A\rh^A
-\lambda^A e^A - \frac{1}{c}\lambda^A_lp^{Al}\Big)\Big]_{,k} +
\hspace{1.4cm}
\\ \nonumber
+\ \Big[\ s^{Ak}
-\chi^AJ^{Am}h^{Ak}_m
-\lambda^Aq^{Ak}
+\lambda^{Ak}p^A
- \lambda^A_l\pi^{Akl}\Big]_{,k}
+
\hspace{1.3cm}
\\ \nonumber
+\kappa^A\Gamma^A+\kappa^A{_{,k}}N^A
+\ \chi^A{_{,k}}J^{Ak}+\chi^AJ^{Ak}{_{,k}}+
\hspace{1.3cm}
\\ \label{T4}
+\ \Big(\lambda^A u^A_l\Big)_{,k}T^{Akl} + \lambda^A u^A_lT^{Akl}{_{,k}}+
\lambda^A{_{l,k}}T^{Akl}+ \lambda^A_{l}T^{Akl}{_{,k}}\ =\
\sigma^A + \varphi^A .\hspace{-.5cm}
\eey
Here, $\sigma^A$ is the entropy production  and $\varphi^A$ the
entropy supply of the $^A$-component. The identity
\R{T4} changes into the entropy production, if $s^A$, $s^{Ak}$
and $\varphi^A$ are specified below.                           
\vspace{.3cm}\newline
The terms of the last row of \R{T4} are scalars which do not contribute to the interior
of the second bracket of \R{T4}. This bracket contains the entropy flux density $s^{Ak}$ and additionally further tensors which all are perpendicular to the 4-velocity $u^{Ak}$. 
The scalars of the last row of \R{T4} do also not contribute to the first row of \R{T4}
because the factor $u^{Ak}$ times a scalar is missing.
\vspace{.3cm}\newline
The first and the third term of the fourth row of \R{T4} become
\bey\nonumber
(\lambda^A u^A_l)_{,k}T^{Akl}\ =\hspace{9.5cm}
\\ \nonumber
=\ \Big(\lambda^A{_{,k}}u^A_l +
\lambda^Au^A{_{l,k}}\Big)
\Big(\frac{1}{c^2}e^Au^{Ak}u^{Al}+
\frac{1}{c}u^{Ak}p^{Al}+\frac{1}{c^2}q^{Ak}u^{Al}+t^{Akl}\Big)=\
\\ \label{T9a3}
=\ \lambda{^A}{_{,k}}u^{Ak}e^A
+\frac{1}{c}\lambda^A u^A{_{l,k}} u^{Ak}p^{Al}
+\lambda^A{_{,k}}q^{Ak}
-\U{p^A\lambda^Au^{Ak}{_{,k}}}
+\lambda^Au^{A}{_{l,k}}\pi^{Akl},\hspace{.1cm}
\eey
and taking \R{cT2} and \R{K15}$_1$ into account\footnote{the signs
$\U{\boxdot}$,
$\overbrace{\boxdot}$,
$\underbrace{\boxdot}$
and $\widetilde{\boxdot}$
mark terms which are related to each other in the sequel}
\bey\nonumber
\lambda^A{_{l,k}}T^{Akl} &=& \lambda^A{_{l,k}} \Big(\frac{1}{c^2}e^Au^{Ak}u^{Al}+
\frac{1}{c}u^{Ak}p^{Al}+\frac{1}{c^2}q^{Ak}u^{Al}+t^{Akl}\Big)\ =
\hspace{1cm}
\\ \nonumber
&=& \lambda^A{_{l,k}}u^{Ak}\frac{1}{c^2}e^Au^{Al}
+\lambda^A{_{l,k}}u^{Ak}\frac{1}{c}p^{Al}
+\frac{1}{c^2}\lambda^A{_{l,k}}q^{Ak}u^{Al}-
\\ \label{Y14}
&&\hspace{5cm}
-\overbrace{\lambda^{A}{_{l,k}}p^Ah^{Akl}} 
+ \lambda^A{_{l,k}}\pi^{Akl}.
\eey
Summing up \R{T9a3} and \R{Y14} results in
\bey\nonumber
(\lambda^A u^A_l)_{,k}T^{Akl}+\lambda^A{_{l,k}}T^{Akl}\ =\hspace{7cm}
\\ \nonumber
=\ \Big(\lambda{^A}{_{,k}}+\lambda^A{_{l,k}}\frac{1}{c^2}u^{Al}\Big)
\Big(q^{Ak}+e^Au^{Ak}\Big)+
\\ \label{Y15}
+\ \Big(\lambda^Au^A{_{l,k}}+\lambda^A{_{l,k}}\Big)
\Big(\pi^{Akl}+u^{Ak}\frac{1}{c}p^{Al}\Big)
-\U{p^A\lambda^Au^{Ak}{_{,k}}}-\overbrace{\lambda^{A}{_{l,k}}p^Ah^{Akl}}.
\vspace{.3cm}\eey
The terms which contain the pressure $p^A$ do not fit in the brackets. Evidently, the
term $\U{p^A\lambda^Au^{Ak}{_{,k}}}$ belongs to the first row of \R{T4}, and the second term can be split by taking \R{K15}$_1$ into account
\bee{T16}
-\overbrace{\lambda^{A}{_{l,k}}p^Ah^{Akl}}\ =\
-\lambda^{Ak}{_{,k}}p^A+ \frac{1}{c^2}\lambda^{A}{_{l,k}}p^Au^{Ak}u^{Al}.
\vspace{.3cm}\ee
In the next section, we now specify $s^A$, $s^{Ak}$ and $\varphi^A$.

\subsection{Exploitation of the entropy identity}
\subsubsection{Entropy density, Gibbs and Gibbs-Duhem equations}

Performing the derivation in the first row of \R{T4} and
inserting the underlined term of \R{T9a3} into the first row of \R{T4}, we
obtain by use of \R{K2}$_1$\footnote{\ $^\td$ is the "component time derivative" $\st{\td}{\boxplus}{^A}:=\boxplus^A{_{,k}}u^{Ak}$}
\bey\nonumber
u^{Ak}{_{,k}}\Big(s^A -\chi^A\rh^Aw^A -\kappa^A\rh^A-\lambda^A e^A - \U{p^A\lambda^A}
- \frac{1}{c}\lambda^A_lp^{Al}\Big)+\hspace{2.1cm}
\\ \nonumber
+\Big(s^A-\chi^A\rh^Aw^A-\kappa^A\rh^A -\lambda^A e^A - \underbrace{p^A\lambda^A}
- \frac{1}{c}\lambda^A_lp^{Al}\Big){^\td}
+\underbrace{(p^A\lambda^A){^\td}}=
\\ \label{T10}
=\ 
\Big[u^{Ak}\Big(s^A-\chi^A\rh^Aw^A -\kappa^A\rh^A -\lambda^A e^A
-p^A\lambda^A
- \frac{1}{c}\lambda^A_lp^{Al}\Big)\Big]_{,k}+\widetilde{(p^A\lambda^A){^\td}}.
\hspace{.8cm}
\vspace{.3cm}\eey
We now define the entropy density $s^A$ according to the second round bracket in \R{T10}
\bey\nonumber
\mbox{\sf Setting II:}\hspace{8cm}
\\ \label{T10a}
s^A\ \st{\td}{=}\ \chi^A\rh^Aw^A +\kappa^A\rh^A +\lambda^A e^A+p^A\lambda^A
+\frac{1}{c}\lambda^A_lp^{Al}.
\eey
The specific entropy with regard to the mass density of the mixture is
\bee{T10b}
c^A\ :=\ \frac{\rh^A}{\rh}:\qquad
\frac{s^A}{\rh}\ =\ (\chi^Aw^A+\kappa^A)c^A
+\lambda^A\frac{e^A}{\rh}+p^A\lambda^A\frac{1}{\rh}
+\frac{1}{c}\lambda^A_l\frac{p^{Al}}{\rh}
\ee
which corresponds to the non-equilibrium state space which does not contain the accessory variables
\C{SCH}
\bee{Y2}
{\sf z}\ =\ \Big(c^A,\frac{1}{\rh},\frac{e^A}{\rh},\frac{p^{Al}}{\rh}\Big).
\ee
These state space variables have the following meaning: $c^A$, $1/\rh$ and $e^A/\rh$
are equilibrium variables, $p^{Al}/\rh$ is a non-equilibrium variable extending the
equilibrium sub-space in the sense of Extended Thermodynamics.
The corresponding Gibbs equation according to \R{T10b} and \R{Y2} is
\bee{Y3}
\Big(\frac{s^A}{\rh}\Big)^\td\ =\ (\chi^Aw^A+\kappa^A)\st{\td}{c}{^A}
+p^A\lambda^A\Big(\frac{1}{\rh}\Big)^\td
+\lambda^A \Big(\frac{e^A}{\rh}\Big)^\td
+\frac{1}{c}\lambda^A_l\Big(\frac{p^{Al}}{\rh}\Big)^\td.
\ee
Differentiation of \R{T10b} results in the Gibbs-Duhem equation by taking \R{Y3} into
account
\byy{Y4}
0 &=& (\chi^Aw^A)^\td c^A+\st{\td}{\kappa}\!{^A}c^A
+(p^A\lambda^A)^\td\frac{1}{\rh}
+\st{\td}{\lambda}\!{^A} \frac{e^A}{\rh} +
\frac{1}{c}\st{\td}{\lambda}{^A_l}\frac{p^{Al}}{\rh},
\\ \label{Y4a}
\quad\longrightarrow\quad
\widetilde{(p^A\lambda^A)^\td} &=& -(\chi^Aw^A)^\td \rh^A
-\st{\td}{\kappa}\!{^A}\rh^A
-\st{\td}{\lambda}\!{^A} e^A
-\frac{1}{c}\st{\td}{\lambda}{^A_l}p^{Al}.
\eey
According to \R{T10} and \R{T10a}, the first row of \R{T4} becomes
$\widetilde{(p^A\lambda^A)^\td}$.

\subsubsection{Entropy flux, supply and production}

According to the second row of \R{T4}, we define the entropy flux density
\bey\nonumber
\mbox{\sf Setting III:}\hspace{8cm}
\\ \label{T7} 
s^{Ak}\ \st{\td}{=}\ \lambda^Aq^{Ak}
+\chi^AJ^{Am}h^{Ak}_m-\lambda^{Ak}p^A+\lambda^A_l\pi^{Akl}.
\hspace{1cm}
\vspace{.3cm}\eey
Consequently, the second row of the entropy identity vanishes and \R{T4} becomes
by taking \R{Y4a}, \R{T7}, \R{Y15} and \R{T16} into account
\bey\nonumber
S^{Ak}{_{,k}}\ \equiv\ 
&-&(\chi^Aw^A)^\td\rh^A-\underbrace{\st{\td}{\kappa}\!{^A}\rh^A}
-\U{\st{\td}{\lambda}\!{^A} e^A}
-\widehat{\frac{1}{c}\st{\td}{\lambda}{^A_l}p^{Al}} +
\\ \nonumber
&+& \kappa^A\Gamma^A
+\underbrace{\kappa^A{_{,k}}N^{Ak}}
+\chi^A{_{,k}}J^{Ak}+\chi^AJ^{Ak}{_{,k}}
+ \lambda^A u^A_lT^{Akl}{_{,k}}+ \lambda^A_{l}T^{Akl}{_{,k}}+
\\ \nonumber
&+& \Big(\U{\lambda{^A}{_{,k}}}+\lambda^A{_{l,k}}\frac{1}{c^2}u^{Al}\Big)
\Big(q^{Ak}+\U{e^Au^{Ak}}\Big)+\hspace{1.8cm}
\\ \nonumber
&+& \Big(\lambda^Au^A{_{l,k}}+\widehat{\lambda^A{_{l,k}}}\Big)
\Big(\pi^{Akl}+\widehat{u^{Ak}\frac{1}{c}p^{Al}}\Big)-\hspace{1.4cm}
\\ \label{T7a}
&&\mbox{}\hspace{1.5cm}
-\lambda^{Ak}{_{,k}}p^A+ \frac{1}{c^2}\lambda^{A}{_{l,k}}p^Au^{Ak}u^{Al}\ =\
\sigma^A + \varphi^A .
\eey
The marked terms cancel each other, and we obtain 
\bey\nonumber
S^{Ak}{_{,k}}\ \equiv\ 
&-&(\chi^Aw^A)^\td\rh^A +\chi^A{_{,k}}J^{Ak}+
\\ \nonumber
&+&  \kappa^A\Gamma^A+\chi^AJ^{Ak}{_{,k}}
+ \lambda^A u^A_lT^{Akl}{_{,k}}+ \lambda^A_{l}T^{Akl}{_{,k}}+
\\ \nonumber
&+& \Big(\lambda{^A}{_{,k}}+\lambda^A{_{l,k}}\frac{1}{c^2}u^{Al}\Big)
q^{Ak}+\lambda^A{_{l,k}}\frac{1}{c^2}u^{Al}e^Au^{Ak}+
\\ \nonumber
&+& \Big(\lambda^Au^A{_{l,k}}+\lambda^A{_{l,k}}\Big)
\pi^{Akl}+\lambda^Au^A{_{l,k}}u^{Ak}\frac{1}{c}p^{Al}-
\\ \label{aT7a}
&&\mbox{}\hspace{1.5cm}
-\lambda^{Ak}{_{,k}}p^A+ \frac{1}{c^2}\lambda^{A}{_{l,k}}p^Au^{Ak}u^{Al}\ =\
\sigma^A + \varphi^A .\hspace{.3cm}
\vspace{.3cm}\eey
We now split the entropy identity \R{aT7a} into the entropy production and the entropy
supply. For this end, we need a criterion to distinguish the production from the supply.
Such a criterion is clear for discrete systems: a local isolation suppresses the entropy
supply but not the entropy production. Here for systems in field formulation, we cannot
apply the local isolation. Instead of that, we define the entropy  supply by terms which
contain the energy and momentum supply, \R{K13} and \R{K14} and beyond
that divergencies. The entropy production is characterized by gradients and fluxes.
Consequently, we define the entropy supply
\bey\nonumber
\mbox{\sf Setting IV:}\hspace{8cm}
\\ \nonumber
\chi^AJ^{Ak}{_{,k}}+
\lambda^A u^A_lT^{Akl}{_{,k}}+\lambda^A_{l}T^{Akl}{_{,k}}
-\lambda^{Ak}{_{,k}}p^A
\ =\hspace{2.7cm}
\\ \label{T8}
=\ \chi^AJ^{Ak}{_{,k}}+
\lambda^A u_l^Ak^{Al} + \lambda^A_mk^{Am}-\lambda^{Ak}{_{,k}}p^A
\ \st{\td}{=}\ \varphi^A.
\eey
Taking \R{L3}$_1$ into account , the first row of \R{aT7a} results in
\bee{T7b}
\chi^A{_{,k}}J^{Ak}-\st{\td}{\chi}\!{^A}\rh^Aw^A-\chi^A\st{\td}{w}{^A}\rh^A\ =\
\chi^A{_{,k}}J^{Am}h^{Ak}_m-\chi^A\st{\td}{w}{^A}\rh^A,
\ee
and according to \R{T8}, the entropy identity changes in the entropy production
\bey\nonumber
\sigma^A &=&
\chi^A{_{,k}}J^{Am}h^{Ak}_m -\chi^A\st{\td}{w}{^A}\rh^A+\kappa^A\Gamma^A+
\\ \nonumber
&&+ \Big(\lambda{^A}{_{,k}}+\lambda^A{_{l,k}}\frac{1}{c^2}u^{Al}\Big)
q^{Ak}+\lambda^A{_{l,k}}\frac{1}{c^2}u^{Al}e^Au^{Ak}+
\\ \nonumber
&&+ \Big(\lambda^Au^A{_{l,k}}+\lambda^A{_{l,k}}\Big)
\pi^{Akl}+\lambda^Au^A{_{l,k}}u^{Ak}\frac{1}{c}p^{Al}-
\\ \label{T8a}
&&\mbox{}\hspace{4cm}
+ \frac{1}{c^2}\lambda^{A}{_{l,k}}p^Au^{Ak}u^{Al}.
\eey
A rearranging results in
\bey\nonumber
\sigma^A &=&
\chi^A{_{,k}}J^{Am}h^{Ak}_m -\chi^A\st{\td}{w}{^A}\rh^A+\kappa^A\Gamma^A+
\\ \nonumber
&&+ \lambda{^A}{_{,k}}q^{Ak}
+\lambda^A{_{l,k}}\frac{1}{c^2}u^{Al}\Big(q^{Ak}+e^Au^{Ak}\Big)+
\\ \nonumber
&&+ \lambda^Au^A{_{l,k}}\Big(\pi^{Akl}+u^{Ak}\frac{1}{c}p^{Al}\Big)
+\lambda^A{_{l,k}}\pi^{Akl}+
\\ \label{T8b}
&&\mbox{}\hspace{4cm}
+ \frac{1}{c^2}\lambda^{A}{_{l,k}}p^Au^{Ak}u^{Al},
\eey
and finally, the entropy production becomes
\bey\nonumber
\sigma^A &=&
\chi^A{_{,k}}J^{Am}h^{Ak}_m +\kappa^A\Gamma^A+
\\ \nonumber
&&+ \lambda{^A}{_{,k}}q^{Ak}+
\lambda^Au^A{_{l,k}}\Big(\pi^{Akl}+u^{Ak}\frac{1}{c}p^{Al}\Big)
-
\\ \label{T7g}
&&-\chi^A\st{\td}{w}{^A}\rh^A+ 
\lambda^A{_{l,k}}\Big[\pi^{Akl}
+\frac{1}{c^2}u^{Al}\Big(q^{Ak}+(e^A+ p^A)u^{Ak}\Big)\Big].
\eey
The first four terms of the entropy production describe the classical four reasons of irreversibility:
diffusion, chemical reactions, heat conduction and internal friction with a modified viscous tensor
according to the chosen state space \R{Y2}. To these irreversible processes belong the following
four accessory variables: $\chi^A$, $\kappa^A$, $\lambda^A$ and $\lambda^A_l$.
The last two terms\footnote{which vanish in equilibrium
and for free 1-component systems as we will see below} are typical for an interacting 1-component
sytem as a part of the mixture. 
\vspace{.3cm}\newline
The expressions of entropy density, production and supply and the entropy flux density
contain these accessory variables which are introduced for formulating the entropy identity
playing up to here the role of place-holders. Their physical meaning is discussed in the next section.

\subsection{Accessory variables\label{ACVA}}

Starting out with \R{T10a}, we have the following equation of physical dimensions
\bee{Z1}
[s^{A}]\ =\ [\lambda^A][e^A].
\ee
Taking \R{T2w}$_1$ and \R{J6} into account, we obtain
\bee{Z2}
\frac{N}{m^2}\frac{1}{K}\ =\ [\lambda^A] \frac{N}{m^2}
\quad\longrightarrow\quad
[\lambda^A]\ =\ \frac{1}{K},
\ee
that means, $\lambda^A$ is a reciprocal temperature belonging to the
$^A$-component. Therefore, we accept the following
\bey\nonumber
\mbox{\sf Setting V:}\hspace{6cm}
\\ \label{Z3}
\lambda^A\ \st{\td}{=}\ \frac{1}{\Theta^A}.\hspace{3cm}
\eey
Here, $\Theta^A$ is the temperature of the $^A$-component\footnote{This
temperature is a non-equilibrium one, the contact temperature \C{MUTEMP,MUTEMP1} which should not be confused with the thermostatic equilibrium temperature $T^A$.}.
\vspace{.3cm}\newline
Starting out with \R{T7} , we have the following equation of physical dimensions
\bee{Z4}
[s^{Ak}]\ =\ [\chi^A][J^{Am}][h_m^{Ak}].
\ee
Taking \R{T2w}$_2$, \R{K8c}$_2$ and \R{J5}$_1$ into account, we obtain 
\bee{Z5}
\frac{N}{ms}\frac{1}{K}\ =\ [\chi^A]\frac{kg}{m^3}\frac{m}{s}1
\quad\longrightarrow\quad
[\chi^A]\ =\ \frac{m^2}{s^2}\frac{1}{K}\ =\ \frac{N}{m^2}\frac{m^3}{kg}\frac{1}{K}\ =\
[\kappa^A],
\ee
and according to the the first row of \R{dT2}, the physical dimensions of $\chi^A$ and $\kappa^A$
are equal. We know from the non-relativistic Gibbs equation that the chemical potentials $\mu^A$
have the physical dimension of the specific energy $e^A/\rh$
\bee{Z5a}
[\mu^A]\ = \frac{[e^A]}{[\rh]}\ =\ \frac{N}{m^2}\frac{m^3}{kg}\ =\
K[\chi^A]\ =\ K[\kappa^A].
\ee
Because $\chi^A$ belongs to diffusion and $\kappa^A$ to chemical reactions,
we make the following choice by taking \R{Z5a} into consideration
\bey\nonumber
\mbox{\sf Setting VI:}\hspace{6cm}
\\ \label{Z7}
\chi^A\ \st{\td}{=}\ \kappa^A\ \st{\td}{=}\ -\lambda^A \mu^A.\hspace{3cm}
\vspace{.3cm}\eey
Starting out with \R{T7} , we have the following equation of physical dimensions
\bee{Z8}
[s^{Ak}]\ =\ [\lambda^{Ak}][p^A].
\ee
Taking \R{T2w}$_2$ and \R{J7} into account, we obtain
\bee{Z9}
\frac{N}{ms}\frac{1}{K}\ =\ [\lambda^{Ak}]\frac{N}{m^2}
\quad\longrightarrow\quad
[\lambda^{Ak}]\ =\ \frac{m}{s}\frac{1}{K}.
\ee
By taking \R{bT2}$_2$ into account, we define
\bey\nonumber
\mbox{\sf Setting VII:}\hspace{6cm}
\\ \label{Z10}
\lambda^{Ak}\ \st{\td}{=}\ \lambda^Au^mh^{Ak}_m.\hspace{3cm}
\vspace{.3cm}\eey
According to \R{Z7}$_1$ and \R{Z10}, we obtain
\byy{Z11}
\chi^A{_{,k}} &=& -\lambda^A{_{,k}}\mu^A - \lambda^A\mu^A{_{,k}},
\\ \label{Z12}
\lambda^A{_{l,k}} &=& (\lambda^Au_mh^{Am}_l)_{,k}\ =\ \lambda^A{_{,k}}u_mh^{Am}_l
+\lambda^A(u_mh^{Am}_l)_{,k}
\eey
Inserting \R{Z11} and \R{Z12} into \R{T7g}, the entropy production change in another expression
\bey\nonumber
\sigma^A &=&
-\lambda^A\mu^A{_{,k}}J^{Am}h^{Ak}_m
-\lambda^A\mu^A\Big(\Gamma^A-\st{\td}{w}{^A}\rh^A\Big)+
\\ \nonumber
&&+ \lambda{^A}{_{,k}}\Big(q^{Ak}-\mu^AJ^{Am}h^{Ak}_m+u_mh^{Am}_l\pi^{Akl}\Big)
\\ \nonumber
&&+
\lambda^Au^A{_{l,k}}\Big(\pi^{Akl}+u^{Ak}\frac{1}{c}p^{Al}\Big)+
\\ \label{Z13}
&&+ 
\lambda^A(u_mh^{Am}_l)_{,k}\Big[\pi^{Akl}
+\frac{1}{c^2}u^{Al}\Big(q^{Ak}+(e^A+ p^A)u^{Ak}\Big)\Big].
\vspace{.3cm}\eey
This expression for the entropy production is composed of following parts:
\begin{itemize}
\item diffusion and chemical reactions with a modified mass production term in the first row of \R{Z13},\vspace{-.3cm}
\item heat conduction with a modified heat flux density in the second row,\vspace{-.3cm}
\item internal friction with a modified viscous tensor in the third row and\vspace{-.3cm}
\item an additional term describing the fact that the $^A$-component is a component of the mixture.
\end{itemize}
Inserting the accessory variables in the expression of entropy density \R{T10a}, of entropy flux
density \R{T7} and of entropy supply \R{T8} we obtain
\byy{Z14}
s^A &=& -\lambda^A\mu^A\Big(w^A +1\Big)\rh^A +\lambda^A e^A+p^A\lambda^A
+\frac{1}{c}\lambda^Au_mh^{Am}_lp^{Al},
\\ \label{Z15}
s^{Ak} &=& \lambda^Aq^{Ak}
-\lambda^A\mu^AJ^{Am}h^{Ak}_m-\lambda^Au^mh^{Ak}_mp^A
+\lambda^Au_mh^{Am}_l\pi^{Akl}
\\ \label{Z16}
\varphi^A &=&
-\lambda^A\mu^AJ^{Ak}{_{,k}}+
\lambda^A u_l^Ak^{Al} + \lambda^Au_ph^{Ap}_mk^{Am}-\Big(\lambda^Au^mh^{Ak}_m\Big)_{,k}p^A.
\vspace{.3cm}\eey
The transition from the interacting $^A$-component to the free 1-component system is considered
in sect.\ref{FC}. All quantities introduced up to here are non-equilibrium ones, because we did not
consider equilibrium conditions up to now. This will be done in the next section.

\subsection{Equilibrium conditions}

Equilibrium is defined by {\em equilibrium conditions} which are divided into
{\em basic} and {\em supplementary} ones \C{MUBO,MUBO1}. The basic equilibrium
conditions are
given by vanishing entropy production, vanishing entropy flux density and vanishing
entropy supply of every $^A$-component\footnote{The sign $\doteq$ stands for a setting which implements an
equilibrium condition.}:
\bee{P23} 
\sigma^{A}_{eq}\ \doteq\ 0\quad\wedge\quad s^{Ak}_{eq}\ \doteq\ 0\quad\wedge\quad
\varphi^A_{eq}\ \doteq\ 0.
\ee
A first supplementary equilibrium condition is the vanishing of all diffusion flux
densities\footnote{Here the definition of equilibrium in composed systems is used: A composed
system is in equilibrium, if all its components are in equilibrium.}.
According to \R{K8c}$_1$, we obtain 
\bee{P23a}
J^{Aeq}_k\ \doteq\ 0\quad\longrightarrow\quad
u^{Aeq}_k\ =\ f^A_{eq}u^{eq}_k\quad\longrightarrow\quad
c^2\ =\ \ f^A_{eq}u^{eq}_k u^{Ak}_{eq}.
\ee
Taking \R{K6b}$_1$ into account, \R{P23a}$_3$ results in
\bee{P23b}
(f^A_{eq})^2\ =\ 1\quad\longrightarrow\quad f^A_{eq}\ =\ \pm 1.
\ee
Consequently, we have to demand beyond \R{P23a}$_1$ the supplementary equilibrium
condition that the mass densities are additive in equilibrium
\bee{P23c}
\rh_{eq}\ \doteq\ \sum_A\rh^A_{eq}\ \longrightarrow\ f^A_{eq}\ =\ 1\ \longrightarrow\
w^A_{eq}\ =\ 0,
\ee
according to \R{K6b}$_2$ and \R{K9a}$_2$.
Taking \R{P23a}$_2$ and \R{Z10} into account, \R{P23c}$_2$ yields
\bee{P23d}
u^{Aeq}_k\ =\ u^{eq}_k\ \longrightarrow\
{\lambda}{^{Ak}_{eq}}\ =\ 0.
\vspace{.3cm}\ee
Further supplementary equilibrium conditions are given by vanishing covariant
time derivatives, except that of the four-velocity:
\bee{P24}
\boxplus^\bullet_{eq}\ \doteq\ 0,\qquad\boxplus\
\neq\ u^l,
\ee
that means $\st{\td}{u}\!{^l_{eq}}$ is in general not zero in equilibrium. Consequently,
the time derivatives of all expressions which contain the 4-velocity must be calculated
separately, as we will see below.  
\vspace{.3cm}\newline
According to \R{P24}$_1$, we obtain
\bee{P25}
\st{\td}{\rh}\!{^A_{eq}}\ =\ 0,\qquad 
\st{\td}{\lambda}{^A_{eq}}\ =\ 0,
\ee
and the (3+1)-components of the energy-momentum tensor, \R{J2} and \R{J3},
satisfy
\bee{aP25}
\st{\td}{e}{^A_{eq}}=0,\quad \st{\td}{p}{^{Al}_{eq}}=0,
\quad \st{\td}{q}{^{Ak}_{eq}}=0,\quad \st{\td}{p}{^A_{eq}}=0,
\quad \st{\td}{\pi}{^{Akl}_{eq}}=0.
\vspace{.3cm}\ee
Starting out with \R{K6b}$_1$, we  have
\bee{P25b}
\st{\td}{f}\!{^A_{eq}}\ =\ \frac{1}{c^2}\Big(\st{\td}{u}{^{Aeq}_m}u^m_{eq}+
{u}{^{Aeq}_m}\st{\td}{u}\!{^m_{eq}}\Big).
\ee
Taking \R{P23d}$_1$ into account, this results in
\bee{bP25}
\st{\td}{f}\!{^A_{eq}}\ =\ 0\quad\longrightarrow\quad
\st{\td}{w}\!{^A_{eq}}\ =\ 0. 
\vspace{.3cm}\ee
According to \R{Z10}, we obtain
\bee{Y16c}
\lambda^A{_{l,k}}\ =\ \Big(\lambda^Au_mh^{Am}_l\Big)_{,k}\ 
=\ \lambda^A_{,k}u_mh^{Am}_l + \lambda^Au_{m,k}h^{Am}_l
+ \lambda^Au_mh^{Am}_{l,k}. 
\ee
In equilibrium, we have according to \R{P23d}$_1$ and \R{K15}
\bee{Y16d}
h^{Am}_{leq}\ =\ h^{m}_{leq}.
\ee
Consequently, the first term of the RHS of \R{Y16c} vanishes in equilibrium, and only the
$\delta^m_l$ remains in the second term. Thus, we obtain
\bee{Y16e}
\lambda^{Aeq}{_{l,k}}\ =\ \lambda^A_{eq}u_{l,k}^{eq}
-\frac{1}{c^2}\lambda^A_{eq}u_m^{eq}
\Big(u^{meq}_{,k}u_l^{eq}+u^m_{eq}u^{eq}_{l,k}\Big)\ =\ 0.
\ee
Consequently, the time derivatives of the accessory variables vanish in equilibrium,
and according to \R{Y3} and \R{Y4}, Gibbs and Gibbs--Duhem equations
are identically satisfied in equilibrium.
\vspace{.3cm}\newline
Taking \R{P23c}$_3$ and \R{Y16d} into account, the entropy density \R{Z14} becomes in
equilibrium
\bee{W1}
s^A_{eq}\ =\ \lambda^A_{eq}\Big(e^A_{eq} + p^A_{eq}- \mu^A_{eq}\rh^A _{eq}\Big).
\ee
This is the usual expression for the entropy density in thermostatics. The energy density and the
pressure are here defined by the (3+1)-decomposition \R{J1} of the energy-momentum tensor.
The chemical potential is as well as the temperature introduced as an accessory variable.
\vspace{.3cm}\newline
Taking \R{P23}$_2$, \R{P23a}$_1$ and \R{Y16d} into account, the entropy flux density \R{Z15}
results in
\bee{Y16a}
0\ =\ q^{Aeq}_k,
\ee
and finally taking \R{P23a}$_1$, \R{Y16d} and \R{Y16e} into account, the entropy supply
\R{Z16} results in
\bee{W2}
0\ =\ u^{Aeq}_lk^{Al}_{eq},
\ee
that means, the power vanishes in equilibrium.
\vspace{.3cm}\newline
Another supplementary equilibrium condition is the vanishing of the production term in
\R{K2}$_2$
\bee{Y15a}
\Gamma^A_{eq}\ \doteq\ 0.
\ee
Thus, we obtain from \R{K2}, \R{P25}$_1$ and \R{Y15a} 
\bee{Y16}
\varrho^A{_{,k}}u^{Ak}+ \varrho^Au^{Ak}{_{,k}}\ =\ \Gamma^A
\quad\longrightarrow\quad u^{Ak}_{eq}{_{,k}}\ =\ 0.
\vspace{.3cm}\ee
The entropy production \R{Z13} has to vanish in equilibrium according to the basic
equilibrium condition \R{P23}$_1$. Taking \R{P23a}$_1$, \R{Y16a},
\R{aP25}$_2$ and  \R{Y16e} into account, we obtain 
\bee{Y16b}
0\ =\ u^{Aeq}{_{l,k}}\pi^{Akl}_{eq}.
\vspace{.3cm}\ee
As demonstrated in this section, the non-equilibrium expressions of the $^A$component in the
mixture for the entropy density, flux, production and supply are compatible with the corresponding
equilibrium expressions. According to \R{P23d}$_1$, the mixture looks like an 1-component
system in equilibrium. But different is,  how to deal with free 1-component systems in
non-equilibrium which are considered in the next section.

\section{Thermodynamics (Free Component)\label{FC}}

A free 1-component system\footnote{that is not a mixture which is treated elsewhere}
can be described by setting equal all component indices of a multi-component system
\bee{O3} 
A, B, C,...,Z\quad\longrightarrow\quad 0,
\ee
and for shortness, we omit this common index 0. Then the basic fields are according to \R{K3}
\bee{O4}
\mbox{rest mass density and 4-velocity:}\hspace{.5cm}\{\rh,u_k\}.
\ee
The equations \R{K5} of Setting I change in identities. According to \R{K6b}$_1$, \R{K8c}$_1$,
\R{K9a}$_2$ and \R{O2}, we have
\bee{O5}
f\ =\ 1,\quad J_k\ =\ 0,\quad w\ =\ 0,\quad T^{kl}{_{,k}}\ =\ k^l.
\ee
The accessory variables become acording to \R{Z3}, \R{Z7} and \R{Z10}
\bee{O6}
\lambda\ =\ \frac{1}{\Theta},\quad \chi\ =\ \kappa\ =\ -\lambda\mu,\quad\lambda^k\ =\ 0.
\ee
The state space \R{Y2} and the entropy density \R{Z14} are as in equilibrium of the
$^A$-component \R{W1}
\bee{O7}
{\sf z}\ =\ \Big(\frac{1}{\rh},\frac{e}{\rh},\frac{p^{l}}{\rh}\Big),\qquad
s\ =\ \lambda\Big(e + p- \mu\rh \Big).
\ee
The entropy flux \R{Z14}, the entropy supply \R{Z16} and the entropy production \R{Z13} are
\bee{O8}
s^k\ =\ \lambda q^k,\quad \varphi\ =\ \lambda u_lk^l,\quad 
\sigma\ =\ \lambda_{,k}q^k +\lambda u_{l,k}\Big(\pi^{kl}+u^k\frac{1}{c}p^l\Big)
-\lambda\mu\Gamma.
\vspace{.3cm}\ee
Another expressions for the entropy production \R{O8}$_3$ is according to \R{J1}
\bee{O10}
\sigma\ =\ \lambda_{,k}q^k +\lambda u_{l,k}
\Big(T^{kl}-\frac{1}{c^2}eu^ku^l-\frac{1}{c^2}q^ku^l+ph^{kl}\Big)-\lambda\mu\Gamma.
\ee
This expression can be transformed by taking
\bee{O11}
-\lambda u_{l,k}\frac{1}{c^2}q^ku^l\ =\ 0
\ee
into account. Consequently, \R{O10} yields
\bee{O12}
\sigma\ =\ \lambda_{,k}q^k +\Big[(\lambda u_l)_{,k}-\lambda_{,k}u_l\Big]
\Big(T^{kl}-\frac{1}{c^2}eu^ku^l+ph^{kl}\Big)-\lambda\mu\Gamma.
\ee
Taking according to \R{J1}
\bee{O13}
u_lT^{kl}\ =\ eu^k+q^k
\ee
into account, the entropy production \R{O12} becomes
\bee{O14}
\sigma\ =\ (\lambda u_l)_{,k}\Big(T^{kl}
-\frac{1}{c^2}eu^ku^l+ph^{kl}\Big)-\lambda\mu\Gamma.
\vspace{.3cm}\ee
We now consider the three terms in the bracket of \R{O14} separately.
Taking \R{J1} into account, \R{O14}
becomes
\bey\nonumber
(\lambda_{,k}u_l+\lambda u_{l,k})\Big(\frac{1}{c^2}eu^{k}u^{l} + 
\frac{1}{c}u^{k}p^{l} +\frac{1}{c^2}q^{k}u^{l} -ph^{kl}+\pi^{kl}\Big)=
\\ \label{O15}
=\ \st{\td}{\lambda}e -\lambda u_l\frac{1}{c}\st{\td}{p}{^l} +\lambda_{,k}q^k -
\lambda p u^k{_{,k}} +\lambda u_{l,k}\pi^{kl},
\\ \label{O16}
-(\lambda_{,k}u_l+\lambda u_{l,k})\frac{1}{c^2}eu^{k}u^{l}\ =\ -\st{\td}{\lambda}e,
\hspace{4.5cm}
\\ \label{O17}
(\lambda_{,k}u_l+\lambda u_{l,k})ph^{kl}\ =\ \lambda p  u^k{_{,k}}.\hspace{5.1cm}
\vspace{.3cm}\eey
The basic equilibrium conditions are \R{P23}$_1$ and \R{Y15a}, and \R{O14} yields
\bee{O18}
0\ =\ (\lambda u_l)_{,k}^{eq}\Big(T^{kl}-\frac{1}{c^2}eu^ku^l+ph^{kl}\Big)^{eq}.
\ee
According to \R{O15} to \R{O17}, this equilibrium condition is satisfied by
\bey\nonumber
(\lambda u_l)_{,k}^{eq}\ \neq\ 0:\hspace{8.6cm}
\\ \label{O19}
(\lambda u_l)_{,k}^{eq}T^{kl}_{eq}\ =\ 0,\quad
(\lambda u_l)_{,k}^{eq}e^{eq}u^k_{eq}u^l_{eq}\ =\ 0,\quad
(\lambda u_l)_{,k}^{eq}p^{eq}h^{kl}_{eq}\ =\ 0.
\eey
Consequently, \R{O18} does not enforce $(\lambda u_l)_{,k}^{eq}=0$. Also the setting
\bee{O20}
T^{kl}_{eq}\ \st{?}{=}\ \frac{1}{c^2}e^{eq}u^k_{eq}u^l_{eq}-p^{eq}h^{kl}_{eq}
\ee
would demand further supplementary equilibrium conditions according to \R{J1}:
\bee{O21}
p^l_{eq}\ \st{?}{=}\ 0,\qquad \pi^{kl}_{eq}\ \st{?}{=}\ 0
\ee
which both are beyond \R{aP25}$_2$ and \R{Y16b}. Therefore we deny \R{O20} and \R{O21},
and $(\lambda u_l)_{,k}^{eq}=0$ is not an equilibrium condition because of the validity of \R{O19}. This property carries over to the Killing relation of the 4-temperature $\lambda u_l$
\bee{O22}
\Big((\lambda u_l)_{,k}+(\lambda u_k)_{,l}\Big)^{eq}\ \st{?}{=}\ 0
\ee
in case of a symmetric energy-momentum tensor. The derivative of the 4-temperature and the Killing
relation are rather conditions for reversible processes because they enforce the entropy production
to be zero without existing equilibrium 
\bee{O23}
(\lambda u_l)_{,k}^{rev}\ =\ 0,\qquad\mbox{or}\ T^{kl} = T^{lk}\!:\ 
\Big((\lambda u_l)_{,k}+(\lambda u_k)_{,l}\Big)^{rev}\ =\ 0.
\ee
A comparison of \R{O19} with \R{O23} points out the difference between equilibrium and reversible
processes.

\section{Summary}

Starting out with the rest mass densities of the components of the multi-component system,
the mass flux densities of the components are defined by introducing their different 4-velocities.
The mixture of the components is characterized by several settings. The first one is the additivity
of the component's mass flux densities to the mass flux density of the mixture. In combination with
the mixture axiom, this setting allows to define mass density and 4-velocity of the mixture and the
diffusion fluxes of the components. The non-symmetric energy-momentum tensor of one component interacting
with the mixture is introduced, and its (3+1)-split together with the component's mass and diffusion
flux densities are generating the entropy identity \C{MUBO}. The exploitation of the entropy identity requires
additional settings: the entropy density, flux and supply. These settings are led by physical
interpretations of entropy density, flux and supply. The entropy production follows from the entropy
identity which restricts possible arbitrariness of defining. 
\vspace{.3cm}\newline
The use of the entropy identity introduces so-called accessory variables. These are temperature,
chemical potential and an additional non-equilibrium variable which characterizes the considered
component to be a part of the mixture. Beside the classical irreversible processes --diffusion,
chemical reactions, heat conduction and friction-- an additional irreversible process appears due to
the embedding of the considered component into the mixture. Differently from the classical case, the
mass production term, the heat flux density and the viscous tensor are modified, so-called effective
quantities.
\vspace{.3cm}\newline
Equilibrium is defined by equilibrium conditions which are divided into
basic and supplementary ones \C{MUBO,MUBO1,BOCHMU}. The basic equilibrium conditions are
given by vanishing entropy production, vanishing entropy flux density and vanishing
entropy supply. Supplementary equilibrium conditions are: vanishing diffusion flux densities,
vanishing component time derivatives\footnote{except that of the 4-velocity}  and vanishing of the
mass production terms. Presupposing these equilibrium conditions, we obtain: all components have
the same 4-velocity, all heat flux densities are zero, the power as well as the divergence of the
4-velocity of each component vanish, and the viscous tensor is perpendicular to the velocity gradient.
\vspace{.3cm}\newline
The corresponding free component is defined by undistinguishable component
indices\footnote{that is not the mixture which is a multi-component system}. This 1-component
system represents the easiest classical case serving as a test, if the interacting component in the
mixture is correctly described. The vanishing of the entropy production in equilibrium is shortly
investigated: the so-called Killing relation of the vector of 4-temperature in the case of a symmetric
energy-momentum tensor is neither a necessary nor
a sufficient condition for equilibrium. Also the statement that materials are perfect in equilibrium
cannot be confirmed.
\vspace{.7cm}\newline
{\bf Ackowledgement} Discussions with  H.-H. v. Borzeszkowski, T. Chrobok and\newline
G.O. Schellstede are gratefully acknowledged.

\end{document}